# LEVERAGING QUANTUM COMPUTING FOR RECOURSE-BASED ENERGY MANAGEMENT UNDER PV GENERATION UNCERTAINTY

**Daniel Müssig***

*Fraunhofer IOSB-AST, Wilhelmsplatz 11, 02826 Görlitz, Germany*

**Mustafa Musab, Markus Wappler, Jörg Lässig**

*Fraunhofer IOSB-AST, Wilhelmsplatz 11, 02826 Görlitz, Germany*

The integration of distributed energy resources, particularly photovoltaic (PV) systems and electric vehicles (EVs), introduces significant uncertainty and complexity into modern energy systems. This paper explores a novel approach to address these challenges by formulating a stochastic optimization problem that models the uncertain nature of PV power generation and the flexibility of bi-directional EV charging. The problem is structured as a two-stage stochastic program with recourse, enabling the system to make optimal day-ahead decisions while incorporating corrective actions in real time based on actual PV output and EV availability. Leveraging the capabilities of quantum computing, we implement and solve the stochastic model using quantum algorithms, demonstrating the potential of quantum-enhanced optimization for high-dimensional and uncertainty-driven energy management problems. Our results indicate that quantum computing can provide efficient and scalable solutions for complex recourse problems in smart grid applications, particularly when integrating variable renewable generation and flexible demand resources.

## 1. Introduction

The rapid growth of renewable energy and electric mobility is transforming today's energy systems. Photovoltaic (PV) panels and electric vehicles (EVs) are key to a sustainable future, yet they also bring new challenges. PV generation depends heavily on weather conditions, which makes it hard to predict. At the same time, EVs can either consume or provide energy to the grid, depending on user behavior and charging flexibility. These uncertainties make it difficult to plan energy use and trading in advance without facing costly imbalances. Stochastic optimization addresses this by formulating the entire problem as a two-stage process with recourse: in the first stage, day-ahead decisions are optimized under forecast uncertainty, and in the second stage, recourse actions are optimized to adapt first stage decisions once actual conditions are realized. This approach enables more reliable and cost-effective energy management but also results in complex mathematical problems. Solving them efficiently with conventional computers is difficult. Quantum computing opens a promising new path. By using the principles of quantum mechanics, quantum algorithms can handle large and uncertain optimization problems in ways that classical methods cannot. In this paper, we present a quantum approach to energy management under PV uncertainty. Our results show how quantum algorithms can support better decision-making in smart grids, demonstrating their potential to manage renewable energy and flexible demand resources at scale.

## 2. Background

Quantum computing builds upon the principles of quantum mechanics to establish a computational paradigm that may surpass classical computing for specific problem classes [1]. The fundamental unit of information is the qubit, which, unlike the classical bit, can exist in a superposition of basis states within a complex Hilbert space. Formally, a qubit is described by the quantum state $|\psi\rangle = \alpha|0\rangle + \beta|1\rangle$, where $|\alpha|^2 + |\beta|^2 = 1$, and $\alpha, \beta \in \mathbb{C}$ are complex probability amplitudes and $|0\rangle, |1\rangle$ are the basis states of the Hilbert space. Central phenomena enabling quantum computation include superposition, entanglement, and interference. In the mathematical formalism, states are represented as vectors in Hilbert space and quantum operations as unitary linear operators. Multi-qubit systems are modeled through tensor products, which naturally capture entanglement by quantum states that cannot be written as a tensor product of individual qubit states.

A key application area for quantum computing is solving optimization problems, particularly combinatorial optimization problems. Early quantum algorithms leverage complex transformations like the Quantum Fourier transform. These usually depend on large circuit depths, which makes them impractical for current noisy intermediate-scale quantum (NISQ) devices. In response, variational quantum algorithms were proposed, utilizing shallow circuits combined with classical optimization as shown in Figure 1.

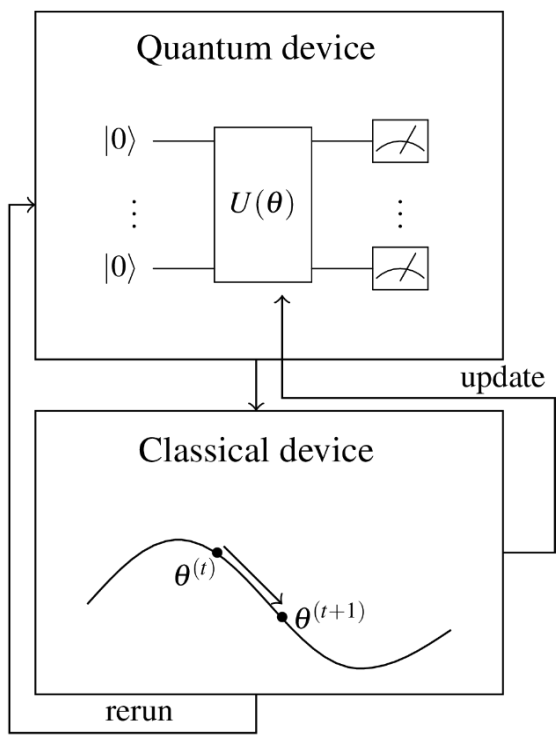

Figure 1: Variational, or hybrid, quantum algorithms combine a parameterized quantum circuit with a classical optimizer: the quantum circuit evaluates the cost function, while the classical computer updates the parameters.

The quantum approximate optimization algorithm (QAOA), introduced by Farhi et al. [2], is tailored for combinatorial optimization problems. It operates by alternately applying a problem Hamiltonian and a mixer Hamiltonian, each parameterized by angles $\gamma_i$ and $\beta_i$, respectively. The depth of the circuit is denoted by the parameter $p$, corresponding to the number of alternations. As $p \to \infty$, QAOA converges towards adiabatic quantum computing; however, even small values of $p$ have been shown to yield strong performance. While $p = 1$ allows for analytical insight via trigonometric forms of the cost function and Fourier analysis of the energy landscape, a higher number of layers potentially achieves quantum advantage but can usually not be examined analytically. For instance, it has been shown that for $p = 8$, QAOA can outperform the Goemans-Williamson approximation algorithm for the Max-Cut problem [3].

QAOA is particularly suitable for quadratic unconstrained binary optimization (QUBO) problems. Once formulated, QUBOs are translated into an Ising Hamiltonian (as shown in Equation (8)) by mapping binary variables $x \in \{0,1\}$ to spin variables $z \in \{-1,1\}$. The corresponding Hamiltonian is then implemented on a quantum circuit using parameterized $RZ$ gates for the external field and $RZZ$ gates for the interaction between two spins. Each layer of the QAOA circuit applies these operators using the same angle γ for all problem terms and β for all mixer terms. The optimization of these parameters is performed classically, with the objective of minimizing the expectation value $\langle\psi(\beta,\gamma)|\ H\_C\ |\ \psi(\beta,\gamma)\ \rangle$. Since we cannot differentiate this function on the quantum computer, gradient-free optimizers are employed. Empirical studies, such as those by Pellow et al. [4], have identified optimizers like ADAM, AMSGrad, and SPSA as effective in various simulation and hardware contexts.

Overall, while quantum advantage remains to be demonstrated, QAOA and its variants constitute a promising avenue for leveraging near-term quantum hardware in the realm of combinatorial optimization.

## 3. Problem Formulation

Stochastic optimization is a promising field for quantum computing, since quantum computers themselves are already stochastic systems. Stochastic optimization is a method to solve optimization problems involving uncertainty. In these problems, some elements, such as inputs, constraints, or objectives, are influenced by random variables or processes. The goal is to find an optimal solution that performs well on average or under most probable scenarios, rather than a deterministic outcome. Within stochastic optimization are many solution approaches defined. The Expected Value Solution assumes the expected values for all random variables and solves it as a deterministic problem. Thus, it is used as a benchmark against stochastic solutions. Another benchmark is the Wait & See solution as it assumes perfect knowledge of the future weighted by its occurrence. A typical stochastic approach is the Here & Now approach. This assumes always a two (or more) staged problem. Decisions must be made before the future realizes. In the next stage one must deal with the consequences of the decisions made in the first stage. A special case of this solution approach is the recourse problem. Here, one can make corrective actions in the later stages to correct for bad decisions in the first stage. Computationally, this leads to a separate minimization problem for each scenario.

To model the stochastic problem, it was decided to formulate it as a recourse problem. The problem states the following: An EV charging provider has a certain number of chargers, an office building and PV installed on top of it. The provider participates in an energy exchange market. Here, it is assumed, that the provider must sell or buy his energy a day ahead. If the energy drawn from or fed into the grid exceeds the day-ahead decisions the provider must sell or buy energy intra-day with worse prices. Further, energy sold by charging EVs will always have a better price then on the energy exchange market. It is assumed, that the company of the office building is the charging provider and that the chargers are not public thus the provider has perfect knowledge on the number of EVs, their arrival time and required amount of energy. Uncertain however, is the PV production, as exact values cannot be predicted day-ahead. But, using weather characteristics such as the cloud cover a range with a probability distribution of the PV production could be predicted.

Following this use case a mathematical formulation of this problem reads:

$$\min z = \sum_t -j_t \cdot 0.25 + E_p Q(j,p) \quad (1)$$

With:

$$Q(j,p) = \min \sum_t (buy_t * 0.4 - sell_t * 0.1) \quad (2)$$

s.t.

$$j_t^{min} \leq j_t \leq j_t^{max} \quad (3)$$

$$j_t - buy_t^i + sell_t^i = p_t^i \quad (4)$$

$$buy_t^i \geq 0 \quad (5)$$

$$sell_t^i \geq 0 \quad (6)$$

$$buy_t^i * sell_t^i = 0 \quad (7)$$

In Equation (1) we want to minimize the costs represented as $z$. For this we sum over $j_t$, which is the energy fed into (positive valued) or drawn (negative valued) from EVs. In the sum we inverse the sign to represent the profit as a negative value. Therefore, the minimization problem maximizes the profit. Here, we sell and buy from and to the EVs with a price of $0.25$. $E_p Q(j,p)$ stands for the expectation value of the function $Q$ (see Equation (2)) under the probability distribution of $p$. In contrast to the general case in our simple example the function $Q$ is not actually a minimization problem, rather it can be calculated directly using the second constraint. As written in the use case, we have either excess PV energy (positive valued) or we have an energy demand (negative valued), which is represented with $p_t$. Note, that for an easier understanding of the formula $p_t$ is defined in Equation (4) and not directly seen in Equation (2). However, through substitution it could be integrated. Equation (3) constraints the maximal energy drawn from the EV as $j_t^{min}$ and fed into the EV as $j_t^{max}$. Through this constraint we can model the physical properties of the charger and the amount of available EVs. In the second stage we have to take corrective actions. We either buy ($buy_t$) or sell ($sell_t$) energy with an intra-day buy price of $0.4$ and an intra-day sell price of $0.1$. Both decision variables are positive valued (Equations (5) and (6)) and at least one of them must be zero (Equation (7)). The superscript $i$ in Equations (4) – (7) represents the scenario, i.e., a specific $p_t$ was chosen from the probability distribution.

## 4.  Quantum Solution Approach

In this section we describe our hybrid quantum solution approach based on QAOA (see Section 2). As written in the background section, classical optimization problems will be usual converted into a QUBO and then transformed into an Ising-Problem, which could be used as problem Hamiltonian for QAOA or quantum annealing. The conversion of a similar problem, although a deterministic one, into a QUBO is shown in one of our other papers [5]. However, to make an uncertainty analysis we must define the Ising-problem here:

$$H_C = -\frac{1}{2}\sum_{i,j} J_{ij} s_i s_j - \sum_i h_i s_i \qquad (8)$$

In Equation (8) $s$ is the spin with $s \in \{+1, -1\}$, which are exactly the eigenvalues of Pauli-$Z$. $J_{ij}$ represents the interactions between two spins $s_i$ and $s_j$. $h_i$ on the other hand, represents the external magnetic field. The terminology stems from the original development of the model, where it was used to describe ferromagnetism in statistical mechanics. It is used in combinatorial optimization on quantum computers and quantum annealers, since it can be directly placed onto them without further conversions or substitutions.

We observe so-called two-body interactions $J_{ij} s_i s_j$, which represent a quadratic term between two decision variables and one-body interactions $h_i s_i$, which represent terms with only one decision variable. $J_{ij}$ and $h_i$ are the constants in our optimization problem. Through a thorough investigation one can find, that $p$ naturally only occurs in $h_i$. Therefore, in our uncertainty analysis we are only interested in $h_i s_i$. As it turns out, this term is represented as the following Quantum-Gate: $RZ(\Phi) = e^{-i\frac{\Phi}{2}Z}$. Since the imaginary component is neither relevant nor problematic and $\frac{1}{2}$ is just a constant, which we can treat classically, we can directly put $h$ in. The $RZ$-Gate is a rotational gate, where consecutive rotations add up, which will be important later. Now that we have seen where our uncertainty appears, we can start implementing the probability distribution into QAOA.

Because of the stochastic nature of quantum computers, we want to encode the probability distribution of $p_t$ into the qubits. Since our values are integers and not binaries, we will use a binary encoding to represent the integers as bitstrings. Here, we will use quantum amplitude encoding to store the probability distribution into a fitting number of qubits (a qubit register). This is done for each timestep $t$. Now, we can use controlled $RZ$-Gates to rotate according to the bitstring stored in the register. If we only consider classical states, then a controlled gate is similar to an $if$. If and only if the control qubit is in state $|1\rangle$ the action on the target qubit is executed. However, for arbitrary quantum states the action is a little bit more complicated, but we can think of it as: in each shot (execution of a circuit) the register is in only one of the possible states from the probability distribution and therefore in a classical state. To apply $p_t$, which is a bit string, we will apply a controlled $RZ$-Gate of $2^i$, where $i \in \{0..n-1\}$ is the ith digit in a binary number counted from least significant bit to the most significant one. A graphical representation of this quantum circuit is given in Figure 2. The rest of the approach follows the typical QAOA-scheme. To summarize the approach, in each QAOA iteration we execute Equation (1) with a specific scenario of the second stage. The scenario is randomly drawn from the given probability distribution.

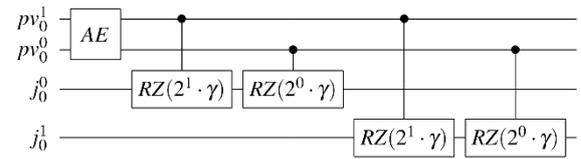

*Figure 2: A quantum circuit consisting of the Amplitude Encoding (AE) and only the RZ-Gates from the problem Hamiltonian of QAOA.*

In the following we want to acknowledge the work of Rotello et al. [6] and their influences on our approach. During the work on our approach the one by Rotello et al. was published. However, with their approach they are calculating the expectation value of the second stage. Our approach on the other hand, calculates the complete Equation (1) as stated above. Although they are not explicitly stating that they are using Quantum Amplitude Encoding in their approach, the exact method used is not that important.

## 5.  Results

To test our model, we have created a very small and simple problem instance, where the solution can be easily verified. Therefore, we use just one timestep and two bits for each decision variable and our uncertain $p$. The probability distribution of $p$ is the following: $\{1: 0.2, 2: 0.5, 3: 0.3\}$. We have also introduced the following simplifications. The constraint in Equation (7) cannot be included in a QUBO directly. However, solutions which violate this constraint will, with the prices given in Equation (1) and (2), naturally be worse than the optimal solution. Further, we have included the constraint in Equation (3) through encoding into our problem instance. We will use two bits for our decision variable $j$, which means, that we can encode the numbers 0 to 3, which will be our $j_{min}$ and $j_{max}$.

We evaluate our approach in the following way using the test problem instance described above. On a Windows Laptop with an Intel(R) Core(TM) Ultra 7 165U 12x2.10 GHz and 32 GB we ran the stochastic QAOA approach with an increasing number of QAOA-layers (from 1 to 100). We also tested different types of parameter initializations. Since QAOA is an approximation algorithm, we run each experiment 50 times to get a good indication of the performance of our approach. As shown in Figure 3 we were able to receive the optimal solution ($j = 2$) with a high probability for 5 layers. However, with increasing number of layers the performance gets worse again. We ran the complete evaluation multiple times and consistently obtained the same result. Regarding the parameter initialization we found that an annealing like schedule with increasing γ and decreasing β parameters worked best. As for the outer classical optimization loop, we found, that COBYLA delivered the best and fastest results.

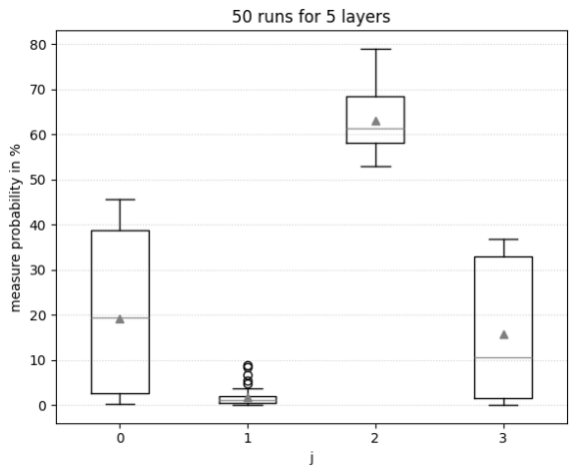

Figure 3: Boxplot of the optimal solution with 5 layers and 50 runs.

The data and source code for running the experiments is available at [7].

## 6. Outlook

This work has introduced a quantum approach to managing energy systems under the uncertainty of photovoltaic generation. By modeling the problem as a two-stage stochastic program with recourse, we demonstrated how day-ahead planning can be combined with intra-day corrective actions. Using a quantum approximate optimization framework, a simplified problem instance showed that quantum computing can find feasible and efficient solutions, with promising results for small system sizes.

While current quantum hardware still imposes limitations, our findings underline the potential of quantum algorithms to address the increasing complexity of energy management in renewable-dominated grids. The ability to naturally incorporate uncertainty and flexibility into the optimization process makes this approach a strong candidate for future applications.

Looking ahead, extending the method to more realistic system sizes, exploring hybrid strategies that combine classical and quantum optimization like Benders Decomposition, and investigating why certain parameter settings perform better than others will be important next steps. Ultimately, this line of research points toward a future where quantum computing contributes to more resilient, cost-efficient, and sustainable energy systems.

## Acknowledgements

Funded by the European Union under Horizon Europe Programme - Grant Agreement 101080086 — NeQST.

Views and opinions expressed are however those of the author(s) only and do not necessarily reflect those of the European Union or European Climate, Infrastructure and Environment Executive Agency (CINEA). Neither the European Union nor the granting authority can be held responsible for them.